\documentclass[floatfix,aps]{revtex4}
\usepackage{graphicx}
\usepackage{amsmath}
\usepackage{amssymb}
\usepackage{bm}

\begin{document}
\title{Single-electron tunneling in the fractional quantum Hall effect regime}
\author{C.W.J. Beenakker and B. Rejaei}
\affiliation{Instituut-Lorentz, Universiteit Leiden, P.O. Box 9506, 2300 RA Leiden, The Netherlands}
\date{June 1993}
\begin{abstract}
A recent mean-field approach to the fractional quantum Hall effect (QHE) is reviewed, with a special emphasis on the application to single-electron tunneling through a quantum dot in a high magnetic field. The theory is based on the adiabatic principle of Greiter and Wilczek, which maps an incompressible state in the integer QHE on the fractional QHE. The single-particle contribution to the addition spectrum is analyzed, for a quantum dot with a parabolic confining potential. The spectrum is shown to be related to the Fock-Darwin spectrum in the integer QHE, upon substitution of the electron charge by the fractional quasiparticle charge. Implications for the periodicity of the Aharonov-Bohm oscillations in the conductance are discussed.\bigskip\\
{\tt Proc.\ NATO Workshop on "Physics of Few-Electron Nanostructures",\\
published in Physica B \textbf{189}, 147 (1993).}
\end{abstract}
\maketitle

\section{Introduction}

Shortly after the discovery \cite{Sco89} and identification \cite{Hou89} of Coulomb-blockade oscillations in semiconductor nanostructures, it became clear that this effect provides a sensitive probe of the ground state properties of a confined, strongly interacting system. Much of the research in the past few years has concentrated on mapping out the energy spectrum in the regime of the {\em integer\/} quantum Hall effect (QHE) \cite{McE91,Joh92,Sta92}. In this regime a conventional mean-field treatment (Hartree or Thomas-Fermi) is sufficient to describe the interaction effects \cite{McE92,Mar92}. These approaches are insufficient for the subtle correlations of the ground state in the {\em fractional\/} QHE. Recently, we have employed the adiabatic principle of Greiter and Wilczek \cite{Gre90} to develop a mean-field theory of the fractional QHE \cite{Rej92}. The many-body correlations are introduced by means of a ficitious vector-potential interaction, which is treated in mean-field. Hence the name {\em vector-mean-field theory}, borrowed from anyon superconductivity \cite{Gro91}.

In this paper we review the work of Ref.\ \cite{Rej92}, with a particular emphasis on the application to single-electron tunneling. Sections 2 and 3 contain a general description of our method. In Sec.\ 4 we specialize to the case of a confined geometry, viz.\ a quantum dot in a two-dimensional electron gas with a parabolic confining potential. We compare the vector-mean-field theory with the exact diagonalization of the hamiltonian for a small system. Sections 5 and 6 deal with the implications of our theory for the periodicity of the conductance oscillations as a function of Fermi energy (Sec.\ 5) and magnetic field (Sec.\ 6). This material was not reported in our previous paper. An issue addressed in these two sections is to what extent the periodicity of conductance oscillations in the fractional QHE can be interpreted in terms of a {\em fractional\/} charge. This issue has been addressed previously \cite{various}, in a different physical context. We conclude in section 7.

\section{Adiabatic mapping}

\begin{figure}[tb]
\centerline{\includegraphics[width=0.3\linewidth,angle=-90]{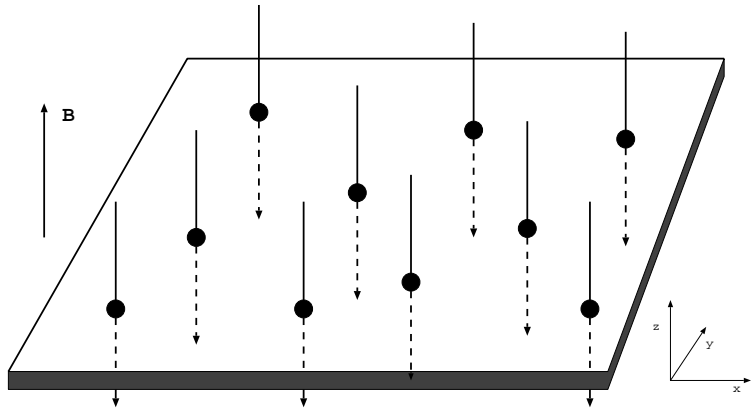}}
\caption{Schematic illustration of the adiabatic mapping. By attaching negative flux tubes to the electrons their density can be reduced adiabatically (the force which increases the electron--electron separation is provided by Faraday's law). Flux tubes containing an even number of flux quanta can be removed instantaneously by a gauge transformation. In this way an initial incompressible state is mapped onto a new incompressible state at lower density. This mapping was proposed by Greiter and Wilczek \cite{Gre90}, to map the integer onto the fractional QHE. In this paper we apply the mean-field approximation of the adiabatic mapping to a confined geometry.
}
\label{schema}
\end{figure}

We consider a two-dimensional electron gas in the $x$-$y$ plane, 
subject to a magnetic field ${\bf B}$ in the $z$-direction. The hamiltonian is
\begin{eqnarray}
{\cal H}_{0} &=& \sum_{i} \frac{1}{2m}
[ {\bf p}_{i} + e {\bf A}({\bf r}_{i})]^{2}
+  \sum_{i<j} u({\bf r}_{i}-{\bf r}_{j}) + \sum_{i} V({\bf r}_{i}) ,\label{H0}
\end{eqnarray}
where ${\bf A}$ is the vector potential associated with ${\bf B}=\nabla \times {\bf A}$, $V$ is an external electrostatic potential, and $u(r)=e^{2}/r$ is the potential of the Coulomb interaction between the electrons. The 
adiabatic principle of Greiter and Wilczek \cite{Gre90} is formulated in terms of a new hamiltonian
\begin{eqnarray}
{\cal H}_{\lambda} &=& \sum_{i} \frac{1}{2m}
[ {\bf p}_{i} + e {\bf A}({\bf r}_{i}) -
e \lambda \sum_{j(\neq i)} {\bf a}({\bf r}_{i}-{\bf r}_{j}) ]^{2}
+  \sum_{i<j} u({\bf r}_{i}-{\bf r}_{j}) + \sum_{i} V({\bf r}_{i}) ,\label{Hlambda}
\end{eqnarray}
which contains an extra vector-potential interaction. The vector 
potential ${\bf a}({\bf r})$ is the field resulting from a flux tube 
in the $z$-direction of strength $h/e$, located at the origin:
\begin{eqnarray}
{\bf a}({\bf r})=\frac{h}{e} \frac{\hat{{\bf z}} \times {\bf r}}{2\pi r^{2}},
\; \nabla \times {\bf a}({\bf r}) = \frac{h}{e} \delta ({\bf r}) \hat{{\bf z}} .\label{adef}
\end{eqnarray}
The hamiltonian ${\cal H}_{\lambda}$ is thus obtained from ${\cal H}_{0}$ by binding a flux tube of strength $-\lambda h/e$ to each of the electrons. The flux tubes point in the direction opposite to the external magnetic field, cf.\ Fig.\ \ref{schema}.

The vector-potential interaction can be eliminated from ${\cal H}_{\lambda}$
by a singular gauge transformation, under which a wave function $\Psi$
transforms as
\begin{eqnarray}
\Psi \longrightarrow \Psi'=\prod_{i,j(\neq i)}
\left( \frac{z_{i}- z_{j}}{\left| z_{i}-z_{j} \right| }
 \right)^{\!\lambda/2}  \Psi,
\end{eqnarray}
where we denote $z_{i}\equiv x_{i}-{\rm i}y_{i}$. For $\lambda=2k$, with
$k$ an integer, this transformation is single-valued and can be written as
\begin{eqnarray}
\Psi \longrightarrow \Psi'=\chi_{2k}\Psi,\;\;\chi_{2k}\equiv
\prod_{i<j} \left( \frac{z_{i}- z_{j}}{\left| z_{i}-z_{j} \right| }
 \right)^{\!2k}.\label{gauge}
\end{eqnarray}
If $\Psi$ is an eigenfunction of ${\cal H}_{2k}$, then $\Psi'$ is an eigenfunction of ${\cal H}_{0}$ with the same eigenvalue.

Let us now start with an eigenstate of ${\cal H}_{0}$, and very 
slowly (adiabatically) switch on the vector-potential interaction
by increasing $\lambda$ from $0$ to $2k$.
After application of the gauge transformation (\ref{gauge}), the initial
eigenstate has evolved into a new, exact eigenstate of the original hamiltonian ${\cal H}_{0}$. The adiabatic principle \cite{Gre90} states that the fractional QHE can be obtained
by adiabatic mapping of the integer QHE. The argument is that an excitation gap is conserved under adiabatic evolution, so that incompressibility of the initial state implies incompressibility of the final state. Suppose that the initial state consists of $N$ electrons fully occupying $p$ Landau levels. This is a uniform and incompressible state of the integer QHE. The initial electron density is $n_{0}=peB/h$, which corresponds to an average of $1/p$ flux quantum per electron. This number is an adiabatic invariant, so that as we attach the negative flux tubes $-\lambda h/e$ to the electrons, their density $n_{\lambda}$ will decrease in such a way that the total average flux $-\lambda h/e+B/n_{\lambda}$ per electron remains equal to $h/pe$. The final density $n_{2k}$ is thus given by
\begin{equation}
-\frac{2kh}{e}+\frac{B}{n_{2k}}=\frac{h}{pe}\Rightarrow n_{2k}=(p^{-1}+2k)^{-1}\frac{eB}{h},
\end{equation}
which corresponds to a fractional filling factor
\begin{equation}
\nu =\frac{p}{2kp+1}.\label{nuJain}
\end{equation}
This is Jain's formula for the hierarchy of filling factors in the fractional QHE \cite{Jai89}. For example, starting from one filled Landau level ($p=1$) and attaching $2k$ negative flux quanta to the electrons, one obtains the fundamental series $\nu=\frac{1}{3},\frac{1}{5},\frac{1}{7},\ldots$. The second level of the hierarchy starts from $p=2$, yielding $\nu=\frac{2}{5},\frac{2}{9},\ldots$. Only filling factors $<\frac{1}{2}$ can be reached by this mapping.

\section{Mean-field approximation}

The adiabatic mapping can in general not be carried out exactly (for an artificial, but exactly soluble model, see Ref.\ \cite{Gre92}). In this section we describe the mean-field approximation proposed in Ref.\ \cite{Rej92}. 
The theory is similar to the vector-mean-field theory \cite{Gro91}
of anyon superconductivity. In this approximation the flux
tubes are smeared out, yielding a fictitious magnetic field ${\bf B}^{{\rm f}}({\bf r})=-\lambda(h/e)n({\bf r}){\bf\hat{z}}$ proportional to the electron density $n$. 
In addition, a fictitious electric field ${\bf E}^{{\rm f}}({\bf r})=\lambda(h/e^{2}){\bf\hat{z}}\times{\bf j}({\bf r})$ proportional to the charge current density ${\bf j}$ is generated
by the motion of the flux tubes bound to the electrons \cite{Rej91}. 
Formally, the mean-field approximation is obtained by 
minimizing the energy functional 
$E^{{\rm MF}}=\langle \Psi^{{\rm MF}}_{\lambda}| {\cal H}_{\lambda}^{\vphantom{\rm MF}}|
\Psi^{{\rm MF}}_{\lambda}\rangle$ in which $|\Psi^{{\rm MF}}_{\lambda}\rangle$ is a 
Slater determinant of single-particle wave functions. 
After dropping the exchange terms (Hartree approximation) 
the single-particle mean-field hamiltonian is found to be
\begin{eqnarray}
\label{HMF}
{\cal H}^{{\rm MF}}_{\lambda}=
 \frac{1}{2m} [ {\bf p}+e{\bf A}({\bf r})-
e\lambda {\bf A}^{{\rm f}}({\bf r}) ]^{2} + e \lambda \Phi^{{\rm f}}({\bf r}) 
+ U({\bf r}) + V({\bf r}) .
\end{eqnarray}
The fictitious vector and scalar potentials ${\bf A}^{{\rm f}}$ and 
$\Phi^{{\rm f}}$ are given by
\begin{eqnarray}
\label{FIC}
{\bf A}^{{\rm f}}({\bf r}) &=& \int\! d{\bf r}' \,{\bf a}({\bf r}-{\bf r}') 
n({\bf r}') , \\
-e\Phi^{{\rm f}}({\bf r}) &=& \int\! d{\bf r}' \,{\bf a}({\bf r}-{\bf r}') 
\cdot {\bf j}({\bf r}'), 
\end{eqnarray}
and the ordinary Hartree interaction potential is 
\begin{eqnarray}
U({\bf r})=\int\! d{\bf r}' \,u({\bf r}-{\bf r}') n({\bf r}') .
\end{eqnarray}
Note that [in view of Eq.\ (\ref{adef})], ${\bf B}^{\rm f}=-\lambda\nabla\times{\bf A}^{\rm f}$ and ${\bf E}^{\rm f}=\lambda\nabla\Phi^{\rm f}$.
The electron and current densities are 
to be determined self-consistently from the relations
\begin{eqnarray}
\label{NJ}
n({\bf r}) &=& \sum_{i=1}^{N} \left| \psi_{\lambda ,i}({\bf r}) \right|^{2} \\
{\bf j}({\bf r}) &=& -\frac{e}{m} \sum_{i=1}^{N}{\rm Re}\,
\psi_{\lambda ,i}^{*}({\bf r}) [ -{\rm i}\hbar \nabla + 
e{\bf A}({\bf r})- e\lambda {\bf A}^{{\rm f}}({\bf r}) ]
\psi_{\lambda ,i}({\bf r}) ,
\end{eqnarray}
where the $\psi_{\lambda ,i}$ are eigenfunctions of 
${\cal H}^{{\rm MF}}_{\lambda}$.

In order to perform the mapping, the set of $N$ equations 
\begin{eqnarray}
\label{SCE}
{\cal H}^{{\rm MF}}_{\lambda} \psi^{\vphantom{\rm MF}}_{\lambda ,i} = \varepsilon^{\vphantom{\rm MF}}_{\lambda ,i}
\psi^{\vphantom{\rm MF}}_{\lambda ,i}
\end{eqnarray}
is to be solved self-consistently as a function of the parameter
$\lambda$, which is varied continuously from 0 to $2k$.  
A further simplification results if the Hartree interaction potential $U$
is also switched on adiabatically by the substitution
$ U \longrightarrow (\lambda/2k) U $. Then,
${\cal H}^{{\rm MF}}_{0}$ describes a system of non-interacting electrons
so that the initial state can be determined exactly.
After application of the gauge transformation (\ref{gauge}), the 
final $N$-electron wave function becomes
\begin{eqnarray}
\label{NPWF}
\Psi^{{\rm MF}}=
\prod_{i<j} \left( \frac{z_{i} - z_{j}}
{\left| z_{i} - z_{j} \right| } \right)^{\! 2k}\frac{1}{N!} 
\sum_{P} (-1)^{P} \psi_{2k,1}({\bf r}_{p_{1}})
\psi_{2k,2}({\bf r}_{p_{2}}) \cdots
\psi_{2k,N}({\bf r}_{p_{N}}),
\end{eqnarray}
where the sum runs over all $N!$ permutations $p_{1},p_{2},\ldots p_{N}$ of $1,2,\ldots N$
and $(-1)^{P}$ is the sign of the permutation.
The interaction energy of the final state is 
given in the mean-field approximation by
\begin{eqnarray}
E_{\rm ee} = \langle \Psi^{{\rm MF}} | \sum_{i<j}u({\bf r}_{i}-{\bf r}_{j}) |
\Psi^{{\rm MF}} \rangle =
 \tfrac{1}{2} \int\! d{\bf r} d{\bf r}'\,
u({\bf r}-{\bf r}') g({\bf r},{\bf r}'),\label{Eee}
\end{eqnarray}
with the pair correlation function $g$ given by
\begin{eqnarray}
g({\bf r},{\bf r}') &=& N(N-1)\int\! d{\bf r}_{3}\cdots d{\bf r}_{N}
| \Psi^{{\rm MF}}({\bf r},{\bf r}',{\bf r}_{3},
\ldots {\bf r}_{N}) |^{2}\nonumber\\
&=&n({\bf r})n({\bf r}')-|D({\bf r},{\bf r}')|^{2},\\
D({\bf r},{\bf r}')&=& \sum_{i=1}^{N} \psi_{2k,i}({\bf r})
\psi_{2k,i}^{*}({\bf r}') .
\end{eqnarray}

We have developed a numerical method to solve these mean-field equations selfconsistently by iteration. In Ref.\ \cite{Rej92} we have compared the mean-field theory with exact results for the ground state energy and excitation gap of an unbounded uniform system. The {\em numerical\/} agreement is not especially good, but still within 10--20\%. What is important is that the {\em qualitative\/} features of the fractional QHE (incompressibility of the ground state, fractional charge and statistics of the excitations) are rigorously reproduced by the vector-mean-field theory \cite{Rej92}. We will not discuss the well-understood unbounded system any further here, but move on directly to a confined geometry.

\section{Quantum dot}

We consider a quantum dot with a 2D parabolic confining potential
\begin{equation}
V(r)=\tfrac{1}{2}m \omega_{0}^{2}(x^{2}+y^{2}),\label{parabola}
\end{equation}
and first summarize some well known facts.
The problem of non-interacting electrons in a uniform external magnetic field $B\hat{\bf z}$ and electrostatic potential (\ref{parabola}) can be solved exactly \cite{Foc28}. The eigenstates of energy and angular momentum in the lowest Landau level are
\begin{equation}
\psi_{l}=(2\pi\ell^{2}2^{l}l!)^{-1/2}(z/\ell)^{l}\exp(-|z|^{2}/4\ell^{2}),\label{psil}
\end{equation}
with the definitions $\ell^{2}\equiv\hbar/m\omega$, $\omega^{2}\equiv\omega_{\rm c}^{2}+4\omega_{0}^{2}$, $\omega_{\rm c}\equiv eB/m$. The integer $l=0,1,2,\ldots$ is the angular momentum quantum number. The energy eigenvalue is
\begin{equation}
\varepsilon_{l}=\tfrac{1}{2}\hbar\omega+\tfrac{1}{2}l\hbar (\omega-\omega_{\rm c}).
\end{equation}
The sum of the single-particle energies of $N$ electrons with total angular momentum $L$ is
\begin{eqnarray}
E_{\rm sp}(N,L)=\tfrac{1}{2}N\hbar\omega+\tfrac{1}{2}L\hbar (\omega-\omega_{\rm c}).\label{Esp}
\end{eqnarray}
Because $\omega_{0}$ enters in the eigenstates (\ref{psil}) only as a scale factor (through $\ell$), the problem of calculating the electron-electron interaction energy can be solved independently of the value of $\omega_{0}$. More precisely, if $E_{\rm ee}(N,L)$ is the  Coulomb interaction energy for $\omega_{0}=0$, then the total energy for $\omega_{0}\neq 0$ is given by
\begin{equation}
E_{\rm tot}(N,L)=(\omega/\omega_{\rm c})^{1/2}E_{\rm ee}(N,L)+E_{\rm sp}(N,L).\label{Etot}
\end{equation}
The groundstate for given $\omega_{0}$ is obtained by choosing the value of $L$ which minimizes $E_{\rm tot}(N,L)$.

\begin{figure}[tb]
\centerline{\includegraphics[width=0.8\linewidth]{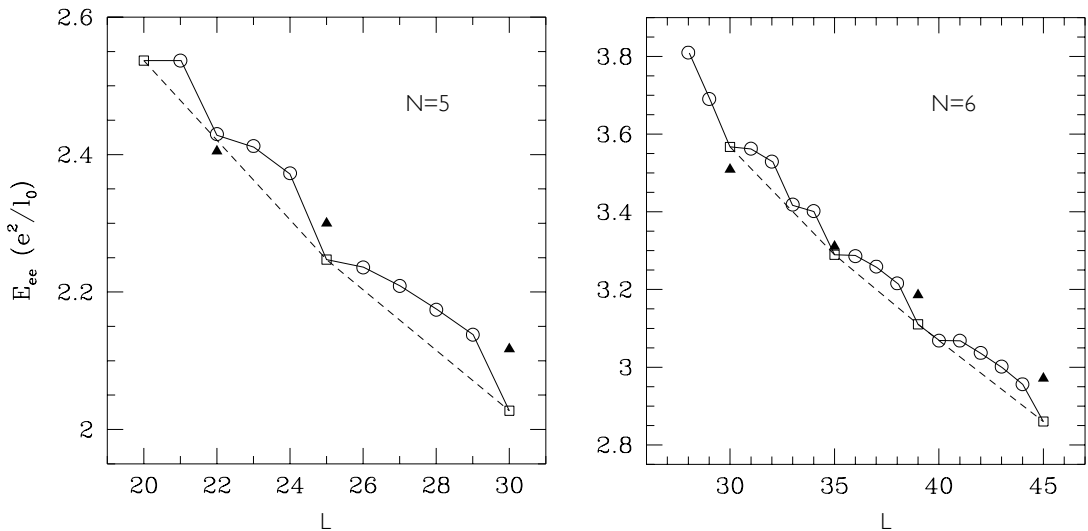}}
\caption{Electron-electron interaction energy of 5 and 6 electrons as a 
function of the angular momentum $L$. The energy is in units of $e^{2}/\ell_{0}$, with $\ell_{0}\equiv (\hbar/eB)^{1/2}$. Triangles follow from the adiabatic
mapping in mean-field approximation. Squares and circles are exact results, squares representing incompressible ground states. Solid lines are a guide to the eye, dashed lines form the Maxwell construction for finding ground states (described in the text). Exact results for $N=6, \: L>39$ could not be obtained because of computational restrictions. The range $L\leq 21\: (N=5)$ and $L\leq 29 \: (N=6)$ can not be reached by adiabatic mapping. (From Ref.\ \protect\cite{Rej92}.)\
}
\label{cusps}
\end{figure}

For small $N$ the interaction energy can be calculated exactly, by diagonalizing the hamiltonian (\ref{H0}) in the space spanned by the lowest Landau level wave functions (\ref{psil}). The technique is described by Trugman and Kivelson \cite{Tru85}. Results for the interaction energy $E_{\rm ee}(N,L)$ as a function of $L$ for 5 and 6 electrons are plotted in Fig.\ \ref{cusps} (open symbols). To determine the ground state value of $L$ one has to minimize $E_{\rm ee}(N,L)+\alpha L$, with $\alpha\equiv\tfrac{1}{2}\hbar(\omega-\omega_{\rm c})(\omega_{\rm c}/\omega)^{1/2}$. This amounts to tilting the plot of $E_{\rm ee}$ versus $L$ with a slope $\alpha$ determined by the strength of the confining potential, and finding the global minimum. The angular momentum values on the convex envelope of the plot (dashed curve in Fig.\ \ref{cusps}) are global minima for some range of $\omega_{0}$. These are the stable incompressible states of the system, at which the interaction energy shows a cusp (squares in Fig.\ \ref{cusps}). Not all cusps are global minima, for example $N=5$, $L=22$ and $N=6$, $L=33$. These cusps are local minima, or ``meta-stable'' incompressible states \cite{Tru85}.

We now turn to the vector-mean-field theory. As initial state of the adiabatic mapping we can choose the incompressible state $|l_{1},l_{2},\ldots l_{N}\rangle=|0,1,\ldots N-1\rangle$ in the lowest Landau level, which has total angular momentum $L_{0}=\tfrac{1}{2}N(N-1)$. This angular momentum is conserved during the adiabatic evolution, during which the electron density is reduced by exchanging mechanical angular momentum for electromagnetic angular momentum (at constant number of electrons in the system). The final gauge transformation (\ref{gauge}) increments the angular momentum by
\begin{eqnarray}
\Delta L=\chi_{2k}^{-1}\sum_{i=1}^{N}\left( z^{\vphantom{*}}_{i}\frac{\partial}{\partial z^{\vphantom{*}}_{i}}-z_{i}^{*}\frac{\partial}{\partial z_{i}^{*}}\right)\chi_{2k}^{\vphantom{-1}}=kN(N-1),\label{DeltaL}
\end{eqnarray}
so that the final angular momentum becomes $L=(k+\tfrac{1}{2})N(N-1)$. This state corresponds to the $\nu=1/(2k+1)$ state in an unbounded system [$p=1$ in Eq.\ (\ref{nuJain})]. 

Just as in the case of an unbounded system, we can start with an incompressible state which occupies $p$ Landau levels. If $N_{n}$ ($n=0,1,2,\ldots$) is the number of electrons in each Landau level, the initial angular momentum eigenvalue is $L_{0}=\tfrac{1}{2}\sum_{n}N_{n}(N_{n}-1-2n)$. Here we have used that the angular momentum eigenvalues in the $n$-th Landau level are $l-n$, with $l=0,1,2,\ldots$. The increment $\Delta L$ is still given by Eq.\ (\ref{DeltaL}), so that the total angular momentum in the final state is \begin{equation}
L=\tfrac{1}{2}\sum_{n=0}^{p-1}N_{n}(N_{n}-1-2n)+kN(N-1).\label{Lfinal}
\end{equation}
To ensure that the initial state is a ground state, each Landau level should be filled up to the same Fermi level. This is achieved by ordering the single-electron energies \cite{Foc28}
\begin{equation}
\varepsilon_{n,l}=\tfrac{1}{2}(n-l)\hbar\omega_{\rm c}+\tfrac{1}{2}(n+l+1)\hbar \omega,\;n,l=0,1,2,\ldots,\label{Fock}
\end{equation}
in ascending order and occupying the $N$ lowest levels. The complete set of incompressible ground states turns out to consist of the set of occupation numbers which satisfy
\begin{eqnarray}
&&N_{0}>N_{1}>\ldots >N_{p-1}>0,\nonumber\\
&&N_{n}=0\;{\rm for}\;n\geq p,\;\sum_{n=0}^{p-1}N_{n}=N.\label{series}
\end{eqnarray}
The occupation numbers of subsequent occupied Landau levels thus have to form a strictly descending series.

In Fig.\ \ref{cusps} the triangular symbols are mean-field interaction energies for $N=5$ and $N=6$. The angular momentum values reached by the adiabatic mapping are dictated by Eqs.\ (\ref{Lfinal}) and (\ref{series}). For example, for $N=5$ the smallest $L$ results from $p=2$, $k=1$, $N_{0}=3$, $N_{1}=2$, yielding $L=22$. For $N=6$, the smallest value of $L$ is obtained by choosing $p=3$, $k=1$, $N_{0}=3$, $N_{1}=2$, $N_{2}=1$, with the result $L=30$. The existence of a smallest value of $L$ corresponds to the restriction $\nu <\tfrac{1}{2}$ in the unbounded system (see Sec.\ 2). It is evident from Fig.\ \ref{cusps} that all the $L$'s reached by adiabatic mapping correspond to a cusp in the exact interaction energy, i.e.\ to a (possibly meta-stable) incompressible state. The adiabatic mapping thus reveals the rule for the ``magic'' angular momentum values of incompressibility.

\begin{figure}[tb]
\centerline{\includegraphics[width=0.4\linewidth]{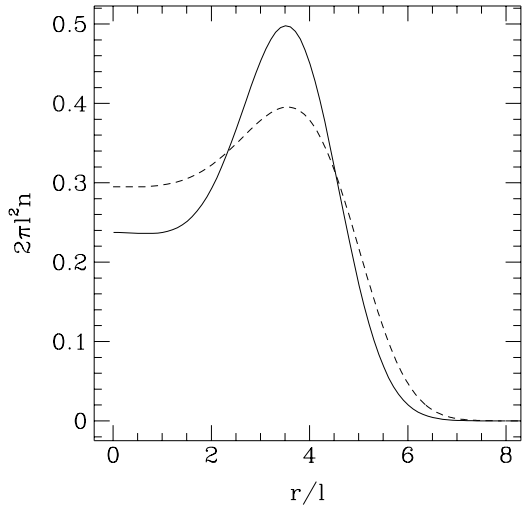}}
\caption{Density profile in a quantum dot with a parabolic confining potential. Comparison of the mean-field theory (solid curve) with the exact result (dotted). The plot is for $N=5$, $L=30$, corresponding to the $\tfrac{1}{3}$ state in an unbounded system. The normalization length $\ell$ is defined in the text [below Eq.\ (\protect\ref{psil})].
}
\label{profile}
\end{figure}

For further comparison between the mean-field theory and the exact diagonalization, we show in Fig.\ \ref{profile} the density profile in the $\tfrac{1}{3}$ state ($N=5$, $L=30$). The agreement is quite reasonable, in particular the curious density peak near the edge (noted in previous exact calculations \cite{Zha85}) is reproduced by the mean-field wave function, albeit with a somewhat smaller amplitude.

\section{Coulomb-blockade oscillations}

Consider the case that the quantum dot is weakly coupled by tunnel barriers to two electron reservoirs, at Fermi energy $E_{\rm F}$. By applying a small voltage difference $V$ between the reservoirs, a current $I$ will flow through the quantum dot. The linear-response conductance $G=\lim_{V\rightarrow 0}I/V$ is an oscillatory function of $E_{\rm F}$. These are the Coulomb-blockade oscillations of single-electron tunneling \cite{Houches}. The periodicity of the conductance oscillations is determined by the $N$-dependence of the ground state energy $U(N)$ of the quantum dot, i.e.\ by its ``addition spectrum''. The condition for a conductance peak is the equality of the chemical potential $\mu(N)\equiv U(N+1)-U(N)$ of the quantum dot and the chemical potential $E_{\rm F}$ of the reservoirs. A conductance peak occurs if $\mu(N)=E_{\rm F}$ for some integer $N$. The spacing of the conductance oscillations as a function of Fermi energy is therefore equal to the spacing $\mu(N+1)-\mu(N)$ of the addition spectrum of the quantum dot.

Let us focus on the analogue of the $\nu=p/(2kp+1)$ state in a quantum dot with a parabolic confining potential. This state results by adiabatic mapping (with the attachment of $2k$ flux tubes) of an incompressible state containing $N\gg 1$ non-interacting electrons distributed equally among $p$ Landau levels. Let $L(N)$ be the ground state angular momentum, computed from Eq.\ (\ref{Lfinal}). The single-particle  (kinetic plus potential energy) contribution $U_{\rm sp}(N)$ to the ground state energy is computed from Eq.\ (\ref{Esp}),\footnote{Equation (\ref{Esp}) gives the kinetic plus potential energy of $N$ electrons with angular momentum $L$ in the {\em lowest\/} Landau level. The final state of the adiabatic mapping needs still to be projected onto the lowest Landau level, since it is not fully in the lowest Landau level for $p>1$. This projection is crucial in order to get the kinetic energy right, but not essential for a calculation of the interaction energy, for which we use the unprojected wave functions.}
\begin{eqnarray}
U_{\rm sp}(N)=\tfrac{1}{2}N\hbar\omega+\tfrac{1}{2}L(N)\hbar (\omega-\omega_{\rm c}).\label{Usp}
\end{eqnarray}
The chemical potential $\mu_{\rm sp}(N)=U_{\rm sp}(N+1)-U_{\rm sp}(N)$ corresponding to $U_{\rm sp}$ depends on which of the occupation numbers of the initial state of the mapping is incremented by one. Suppose that $N_{n}\rightarrow N_{n}+1$. The chemical potential for this transition is
\begin{equation}
\mu_{\rm sp}(N,N_{n})=\tfrac{1}{2}\hbar\omega+\tfrac{1}{2}(2kN+N_{n}-n)\hbar (\omega-\omega_{\rm c}).\label{musp}
\end{equation}
In order to remain in the $p/(2kp+1)$ state, the electrons which are added to the quantum dot have to be distributed equally among the $p$ Landau levels in the initial state of the mapping. This implies that if the transition $N\rightarrow N+1$ was associated with $N_{n}\rightarrow N_{n}+1$, then $N+p\rightarrow N+p+1$ is generically associated with $N_{n}+1\rightarrow N_{n}+2$. The chemical potential difference between these two transitions is
\begin{equation}
\delta\mu_{\rm sp}\equiv\mu_{\rm sp}(N+p,N_{n}+1)-\mu_{\rm sp}(N,N_{n})=\tfrac{1}{2}(2kp+1)\hbar (\omega-\omega_{\rm c}),\label{deltamusp}
\end{equation}
independent of $N$ and $n$. 

We conclude that the kinetic plus potential energy contribution to the addition spectrum of a quantum dot in the $p/(2kp+1)$ state consists of $p$ interwoven series of {\em equidistant\/} levels, each series having the {\em same\/} fundamental spacing $\delta\mu_{\rm sp}$. For $k=0$ we recover the spacing $\varepsilon_{n,l+1}-\varepsilon_{n,l}=\tfrac{1}{2}\hbar (\omega-\omega_{\rm c})$ of the single-electron levels (\ref{Fock}) within a given Landau level, which is independent of $p$. Our Eq.\ (\ref{deltamusp}) generalizes this old result of Fock and Darwin \cite{Foc28} to the fractional QHE. We emphasize that Eq.\ (\ref{deltamusp}) is directly a consequence of the adiabatic mapping described in Sec.\ 4, and does not rely on the mean-field approximation. We can write $\delta\mu_{\rm sp}$ in a more suggestive way in the high-field limit $\omega_{\rm c}\gg\omega_{0}$, when $\omega-\omega_{\rm c}\approx 2\omega_{0}^{2}/\omega_{\rm c}$. Eq.\ (\ref{deltamusp}) then takes the form
\begin{equation}
\delta\mu_{\rm sp}\approx\frac{\hbar m\omega_{0}^{2}}{e^{\ast}B},\;e^{\ast}\equiv\frac{e}{2kp+1}.\label{ester}
\end{equation}
The fundamental spacing in the single-particle addition spectrum in the fractional QHE is therefore obtained from that in the integer QHE by replacing the bare electron charge $e$ by a reduced charge $e^{\ast}$. This reduced charge is recognized as the fractional charge of the quasiparticle excitations in the $p/(2kp+1)$ state \cite{QHE}, although here it appears as a ground state property (and only in the limit $\omega_{\rm c}\gg\omega_{0}$).

So far we have considered only the single-particle contribution to the chemical potential. For a model system with short-range interactions, this is the dominant contribution. Coulomb interactions contribute an amount of order $e^{2}/C$ to the level spacing in the addition spectrum, with the capacitance $C$ of the order of the linear dimension of the quantum dot \cite{Houches}. In typical nanostructures, this charging energy dominates the level spacing, and it would be difficult to extract the $1/e^{\ast}$-dependence from the background $e^{2}/C$ in the periodicity of the conductance oscillations as a function of Fermi energy (or gate voltage). 

We conclude this section by briefly discussing the {\em amplitude\/} of the conductance oscillations. It has been shown by Wen \cite{Wen90} and by Kinaret et al.\ \cite{Kin92} that the (thermally broadenend) conductance peaks in the $1/(2k+1)$ state are suppressed algebraically in the large-$N$ limit. This suppression is referred to as an ``orthogonality catastrophe'', because its origin is the orthogonality of the ground state $|\Psi_{N+1}\rangle$ for $N+1$ electrons to the state $c^{\dagger}|\Psi_{N}\rangle$ obtained when an electron tunnels into the quantum dot containing $N$ electrons. More precisely, the tunneling probability \cite{Mei92} is proportional to $|{\cal M}|^{2}= | \langle \Psi_{N+1}| c^{\dagger}_{\Delta L} | \Psi_{N}\rangle |^{2}$, where $\Psi_{N}$ 
is the $N$-electron ground state and the 
operator $c^{\dagger}_{\Delta L}$ creates an electron in the lowest Landau 
level with wave function $\psi_{\Delta L}$ and angular momentum 
$\Delta L=L(N+1)-L(N)$. Wen and Kinaret et al.\ find that $|{\cal M}|^{2}$ vanishes as $N^{-k}$ when $N \rightarrow \infty$. In the integer QHE, the overlap is unity regardless of $N$. We have investigated whether the vector-mean-field theory can reproduce the orthogonality catastrophe. The calculation is reported in Ref.\ \cite{Rej92}. The result for the $\tfrac{1}{3}$ state ($k=1$) is that $|{\cal M}|^{2}\propto N^{-2}$ for $N\gg 1$. (We have not been able to find a general formula for arbitrary $k$.) 
We conclude that the vector-mean-field theory reproduces the algebraic decay of the tunneling matrix element for large $N$, but with the wrong value of the exponent. In the present context, the orthogonality catastrophe originates from the correlations created by the gauge transformation (\ref{gauge}), required to remove the fictitious vector potential from the hamiltonian (\ref{Hlambda}).

\section{Aharonov-Bohm oscillations}
In the previous section we considered the oscillations of the conductance as a function of Fermi energy (or gate voltage). In the present section we discuss the oscillations in the conductance as a function of magnetic field. Is it possible to identify these magnetoconductance oscillations as $h/e^{\ast}$-Aharonov-Bohm oscillations? A similar question has been addressed in Ref.\ \cite{various}, for different physical systems (a Hall bar or annulus, rather than a quantum dot). We note that in a quantum dot (a singly-connected geometry) the periodicity of the magnetoconductance oscillations in not constrained by gauge invariance. In a ring, in contrast, gauge invariance requires an $h/e$-periodicity of the oscillations, regardless of interactions. The transition from dot to ring has been discussed for the integer quantum Hall effect in Ref.\ \cite{Bee91}.

There is an artificial model which can be solved exactly, and which permits such an identification. This is the model of a hard-core interaction, $u(r)\propto(\nabla^{2})^{2k-1}\delta({\bf r})$. In this model the $1/(2k+1)$ Laughlin state is the exact ground state (for some range of $\omega_{0}$), with {\em vanishing\/} interaction energy \cite{Tru85}. The single-particle energy (\ref{Usp}) is then the whole contribution to the ground state energy. One therefore has [using $L(N)=(k+\tfrac{1}{2})N(N-1)$]
\begin{equation}
U(N)=\tfrac{1}{2}N\hbar\omega+\tfrac{1}{2}(k+\tfrac{1}{2})N(N-1)\hbar(\omega-\omega_{\rm c}).\label{Ushortrange}
\end{equation}
The chemical potential $\mu(N)=U(N+1)-U(N)$ becomes (for $N\gg 1$ and $\omega_{\rm c}\gg\omega_{0}$)
\begin{equation}
\mu(N)=\tfrac{1}{2}\hbar\omega_{\rm c}+(2k+1)N\hbar\omega_{0}^{2}/\omega_{\rm c}.\label{mushortrange}
\end{equation}
A conductance peak occurs when $\mu(N)=E_{\rm F}$. To determine the spacing of the peaks as a function of magnetic field, we have to specify how the Fermi energy $E_{\rm F}$ in the reservoir varies with $B$. The precise dependence is not crucial for our argument. A convenient choice is $E_{\rm F}=V_{0}+\tfrac{1}{2}\hbar\omega_{\rm c}$, with $V_{0}$ an arbitrary conduction band offset. The magnetic field $B_{N}$ of the $N$-th conductance peak is then given by
\begin{equation}
B_{N}=(2k+1)N\frac{\hbar m\omega_{0}^{2}}{eV_{0}},\label{BN}
\end{equation}
and hence the spacing of the peaks is
\begin{equation}
\Delta B=\frac{\hbar m\omega_{0}^{2}}{e^{\ast}V_{0}},\label{DeltaB}
\end{equation}
with $e^{\ast}\equiv e/(2k+1)$ the fractional quasiparticle charge in the $1/(2k+1)$ state. For this hard-core interaction model, the periodicity of the Aharonov-Bohm oscillations in the fractional QHE is thus obtained from that in the integer QHE by the replacement $e\rightarrow e^{\ast}$.

For Coulomb interactions, the chemical potential contains an extra contribution of order $Ne^{2}/C$. The spacing $\Delta B$ of the magnetoconductance oscillations is then increased by a factor $1+e^{2}/C\delta\mu_{\rm sp}$, with $\delta\mu_{\rm sp}=(2k+1)\hbar\omega_{0}^{2}/\omega_{\rm c}$ the single-particle spacing of the addition spectrum. This factor spoils the $1/e^{\ast}$ dependence of the periodicity. For $e^{2}/C\gg\delta\mu_{\rm sp}$ the periodicity of the oscillations is lost altogether. This is the Coulomb blockade of the Aharonov-Bohm effect \cite{Bee91}.

The situation is qualitatively different in the $p/(2kp+1)$ states with $p\geq 2$. In that case the chemical potential depends non-monotonically on $B$ for a constant $N$, because the population $N_{n}$, $n=0,1,\ldots (p-1)$, of the Landau levels in the initial state of the adiabatic mapping varies with $B$ (at constant $N=\sum_{n}N_{n}$). Each transition $N_{n}\rightarrow N_{n}\pm 1$ shows up as a change in slope of $\mu(N)$ as a function of $B$. The equation $\mu(N)=E_{\rm F}$ can therefore have more than a single solution for a given $N$, and hence a series of conductance peaks can occur as $B$ is varied, without incremental charging of the quantum dot. A similar scenario happens in the integer QHE, when more than a single Landau level is occupied \cite{McE91}. The adiabatic mapping thus predicts that the Coulomb blockade suppresses the Aharonov-Bohm oscillations in a quantum dot of small capacitance, if the dot is in the $1/(2k+1)$ state, but not in the $p/(2kp+1)$ states with $p\geq 2$.

\section{Conclusions}

We have shown how the adiabatic mapping of Greiter and Wilczek \cite{Gre90} can form the basis of a mean-field theory of the fractional QHE in a quantum dot with a parabolic confining potential. The angular momentum values obtained by adiabatic mapping of an incompressible ground state in the integer QHE reproduce the ``magic'' values which follow from exact diagonalization of the hamiltonian for a small number $N$ of electrons in the dot. The non-Fermi-liquid nature of the mean-field ground state is illustrated by the algebraic suppression of the probability for resonant tunneling through the dot in the limit $N\rightarrow\infty$ (the orthogonality catastrophe of Wen and Kinaret et al.\ \cite{Wen90,Kin92}).

The vector-mean-field theory provides insight into the addition spectrum of a quantum dot in the fractional QHE, which is the quantity measured by the Coulomb-blockade oscillations in the conductance as a function of Fermi energy. In this paper we have focused on the single-particle (kinetic and potential energy) contribution to the addition spectrum. This is expected to be the dominant contribution for short-range interactions. We have shown that the single-particle addition spectrum in the $\nu=p/(2kp+1)$ state consists of $p$ interwoven series of equidistant levels, similar to the Fock-Darwin single-particle spectrum for $p$ filled Landau levels \cite{Foc28}. The level spacing is renormalized by the substitution $e\rightarrow e^{\ast}$, with $e^{\ast}\equiv e/(2kp+1)$ the fractional quasiparticle charge.

A similar fractional-charge interpretation can be given to the $h/e^{\ast}$-Aharonov-Bohm type oscillations in the conductance as a function of magnetic field. An exactly solvable model was considered, involving a hard-core interaction \cite{Tru85}. For realistic long-range interactions, the charging energy $e^{2}/C$ spoils the simple $1/e^{\ast}$ dependence of the periodicity. We predicted, from the adiabatic mapping, that the periodic Aharonov-Bohm oscillations in a quantum dot with small capacitance are suppressed for filling factors $\tfrac{1}{3},\tfrac{1}{5},\tfrac{1}{7},\ldots$, but not for higher levels of the hierarchy. This prediction should be amenable to experimental verification.

\acknowledgments
This research was supported in part by the ``Ne\-der\-land\-se or\-ga\-ni\-sa\-tie voor We\-ten\-schap\-pe\-lijk On\-der\-zoek'' (NWO) via the ``Stich\-ting voor Fun\-da\-men\-teel On\-der\-zoek der Ma\-te\-rie'' (FOM).

\end{document}